\documentclass[11pt]{amsart}
\usepackage{amsfonts,amssymb, amscd, latexsym, graphicx, psfrag, color,float}
\usepackage{bbm}

\usepackage[all]{xy}

\usepackage{booktabs,enumerate}
\usepackage{multirow}
\usepackage{mathrsfs}

\usepackage{tikz}
\usetikzlibrary{matrix}
\usetikzlibrary{shapes.geometric,decorations.pathreplacing}

\usepackage{tikz-cd}

\usepackage[margin=1in,marginparwidth=0.8in, marginparsep=0.1in]{geometry}

\usepackage[bookmarks=true, bookmarksopen=true,%
bookmarksdepth=3,bookmarksopenlevel=2,%
colorlinks=true,%
linkcolor=blue,%
citecolor=blue,%
filecolor=blue,%
menucolor=blue,%
urlcolor=blue]{hyperref}

\theoremstyle{definition}

\newcommand{\bC}{\mathbb{C}}

\newcommand{\bP}{\mathbb{P}}

\newcommand{\bR}{\mathbb{R}}
\newcommand{\bZ}{\mathbb{Z}}

\newcommand{\cM}{\mathscr{M}}
\newcommand{\cN}{\mathcal{N}}
\newcommand{\cP}{\mathscr{P}}
\newcommand{\cQ}{\mathcal{Q}}
\newcommand{\cR}{\mathcal{R}}

\newcommand{\op}{\operatorname}

\renewcommand{\log}{{\op{log}}}

\newcommand{\Li}{\mathrm{Li}}

 \numberwithin{equation}{subsection}
\numberwithin{figure}{subsection}

\usepackage{tikz}

\numberwithin{equation}{section} 

\title[Wavefunctions for a class of branes in three-space]{Wavefunctions for a class of branes in three-space} 
\author[Eric Zaslow]{Eric Zaslow\\ \vskip0.1in
Department of Mathematics, Northwestern University}

\begin{document}

\begin{abstract}

{Wavefunctions are proposed for some Lagrangian branes in ${\bC^3}$.  The branes are asymptotic to Legendrian surfaces of genus $g$.
The expansion of these wavefunctions in appropriate coordinates conjecturally encodes all-genus open Gromov-Witten invariants, i.e. the free
energy of the topological open string.  

This paper is written in physics language, but tries to welcome mathematicians.  Most results stem from joint mathematical works \cite{SZ,TZ}
with Linhui Shen and David Treumann.}

\end{abstract}

\maketitle


\vskip-.2in
\setcounter{tocdepth}{1}
{\small{\tableofcontents}}
\vskip-.2in
This short note is written to relate several mathematical and physical works and to explain how some of the author's recent collaborations fit in.
Much is known to experts; this is a Festivus for the rest of us.
The discussion is meant to be informal and readable, with details to appear in \cite{SZ}. 
Many footnotes are included to provide some clarification, with mathematical readers in mind.
These footnotes are independent remarks,
not notationally consistent with one another.

\section{Branes and Wavefunctions}
\label{sec:ias}

Witten \cite{W} interpreted the holomorphic anomaly equations of \cite{BCOV} as defining wavefunctions for the toplogical string.\footnote{The topological
string on a Calabi-Yau three-fold $X$ is defined by intersection theory on the space of (stable) holomorphic maps from Riemann surfaces $\Sigma$ of any genus to $X$.
These spaces represent worldsheet instantons for the quantum field theory whose fields are maps $\varphi:  \Sigma \to X$; 
instantons solve $\overline{\partial}\varphi = 0.$  The topological twist ensures that the path integral \emph{only} receives contributions from such maps.  Mathematically,
this theory is Gromov-Witten theory.  The topological string amplitude is a generating function, summing over all genera.}
The wavefunction defines a state in the Hilbert space of
the quantization of the moduli space of complex structures of a Calabi-Yau.\footnote{If $X$ is a Calabi-Yau three-fold,
the period integrals define a map from the moduli space $\cM$ of Calabi-Yau structures of $X$ (complex structure plus choice of Calabi-Yau form)
to the period domain $H^3(X,\bC).$  The point $X\in \cM$ defines
a Lagrangian subspace of the symplectic vector space $H^3(X,\bC)$, with wedge pairing providing the symplectic structure. 
Naively, we quantize a symplectic vector space by choosing a Lagrangian subspace (positions) and writing wavefunctions
as functions of positions.  But then a different choice would yield a different Hilbert space of functions.  So less naively, we want a (projectively) flat connection of the bundle
of such Hilbert spaces over the space of
all Lagrangian.  Then \emph{the} Hilbert space is defined as the global sections.  Pulling this flat connection back to $\cM$ and writing the equation for flatness,
Witten finds it to be the holomorphic anomaly equations of BCOV \cite{BCOV}.}
In a sense, this state can be thought of as the maximally symmetric
vacuum of the theory, with other states
corresponding to the presence of branes, created by ``brane-creation'' operators.  We are interested in these brane states.

Evidence for the wavefunction interpretation comes from several rich examples.  In
\cite{ADKMV}, the wavefunction $\Psi$ for the Aganagic-Vafa brane in the $\bC^3$ geometry --- or equivalently the mirror brane $\{x=const,\, y=const,\, w = 0\}$ in the mirror geometry
$\{wz = x + y - 1\} \subset \bC^2 \times (\bC^*)^2$ --- was shown to satisfy a
quantum equation $H\Psi = 0,$ where $H$ was an operator defined by quantizing the function $x+y-1$ defining the moduli space of the brane.\footnote{Ooguri-Vafa \cite{OV}
studied disk instantons Lagrangian corresponding to knot conormal after the conifold transition,
and computed the unknot conormal in particular.  Aganagic-Vafa computed disk instantons via mirror symmetry for versions of the unknot
conormal in various phases and geometries including and generalizing the resolved conifold.  These results were generalized in \cite{AKV}, where the
role of framing was clarified.}
For knot conormals which have passed through the conifold transition, these functions $H$ have been studied as a quantization of the A-polynomials of the knot
\cite{AV2,GS},\footnote{The definition of the A-polynomial of a knot is essentially as follows.  The boundary of a knot complement is a torus,
and after a framing we identify the rank-one flat bundles with the algebraic torus $\bC^*\times \bC^*.$  This space is symplectic, deriving its structure
from the intersection form.  The $SL_2(\bC)$ character variety of the knot complement maps to this algebraic torus by restriction to the boundary,
and the image is a one-dimensional subvariety defined by a polynomial in two variables, the A-polynomial $A$ of the knot.  The symplectic structure defines
a quantization, with $A$ quantized to an operator $\hat A.$  The quantization of the ideal defined by the A-polynomial is a state in a representation of the quantum torus which is annihilated $\hat A.$} with the example of Aganagic-Vafa being the case of the unknot (first in the resolved conifold \cite{AV}, then in $\bC^3$ \cite{AKV}).
In \cite{AENV} these wavefunctions were obtained for knot and link conormals, and the wavefunction interpretation was referred to as the D-model.
For rigid Lagrangian branes which cannot deform within a fixed Calabi-Yau but do deform along with variations of its complex structure, Walcher-Neitzke \cite{NW} found that
Walcher's open-string version of the BCOV equations were obeyed after a shift of a closed string coupling.\footnote{Such a shift can be effected by a
brane-creation
operator, just as Taylor series tells us that translation by $a$ is effected by  $e^{a\partial}.$}

In this paper we discuss some other Lagrangian branes in $\bC^3$, following \cite{TZ,SZ}. 
They asymptote to genus-$g$ Legendrian surfaces at infinity, although they are not exact.
There is a $g$-dimensional moduli space of such branes.  They are mirror to some of the 
branes considered in 
\cite{ADKMV} and are natural generalizations of the $g=1$ case studied by Aganagic-Vafa.
They are closely related to the conormal of $g$ unlinked unknots through the conifold transition (i.e. $g$ disjoint solid tori).

\section{Ooguri-Vafa Integrality}

We first review the scenario without branes as studied by Gopakumar and Vafa \cite{GV}.
Consider Type-IIA string theory on $\bR^{3,1}\times X,$ where $X$ is a Calabi-Yau threefold.
A stunning discovery of \cite{AGNT,BCOV} was the identification of certain ``F-terms'' in the four-dimensional effective theory
that could be computed with \emph{topological} strings.\footnote{Recall that if $X$ is small, any variation in the $X$ directions
is energetically costly, so at mundane energy levels the fields are ``constant'' in the $X$ directions and effectively only depend on four dimensions.
This is called the ``effective theory,'' and for Calabi-Yau compactification of IIA string theory it is a quantum field theory with $\cN=2$ supersymmetry.
Such theories contain ``F-terms'' with parameters which do not mix with other terms, meaning they may be computed in favorable limits.}
Gopakumar and Vafa \cite{GV} realized that these contributions could be computed at strong (string) coupling, meaning via M-theory,\footnote{Witten's M-theory \cite{W2}
is a description of string theory at strong coupling as 11-dimensional supergravity.} where strings and D2-branes are unifed by M2-branes
(either wrapping the M-theory circle or not, respectively).  Their calculation computed the contribution to these F-terms from such BPS branes, and
since the number of such are integers, this resulted in an expression of the topological string generating function in terms of integers.
The relation between all-genus Gromov-Witten invariants
and integer counts of BPS states is now the celebrated Gopakumar-Vafa integrality conjecture, a vast generalization of the so-called Aspinwall-Morisson formula
which counts $d-$fold covers of a rational curve with contribution $1/d^3.$\footnote{
The preprint \cite{IP} announces a proof.}

Ooguri and Vafa \cite{OV} studied a similar set-up, but now with a either a D4-brane filling two spacetime dimensions cross a Lagrangian three-cycle $L\subset X$,
or a
D6-brane filling spacetime cross $L$.\footnote{This
option is only available if $X$ is noncompact, which is true in our case since $X = \bC^3.$}  In the latter case, the four-dimensional
theory now has $\cN=1$ supersymmetry.  
Similarly to the case without branes above, it has terms in its effective four-dimensional $\cN=1$ theory
which are captured by open-string Gromov-Witten theory and expressable via integer invariants. 
In the M-theory version of the D4-brane set-up,
an M5-brane fills three dimensions of spacetime, and Ooguri-Vafa organize terms of the Lagrangian in terms of integer counts of BPS states in 
the three-dimensional effective $\cN=2$ theory \cite{OV,DW}.  (Ooguri-Vafa also give a four-dimensionsal argument involving BPS domain walls of an
$\cN=1$ theory with central charge \cite{OV}.)

%
%
%

For example, one notable term is the superpotential $W,$ and Ooguri-Vafa integrality counts $d$-fold covers of a disk with contribution $1/d^2.$ 
The sum $\sum_{n=1}^\infty x^d/d^2$ is the dilogarithm $\Li_2(x)$.

It will be important for us to look more closely at the Ooguri-Vafa expansion.  Equation (4.4) of \cite{OV} writes the free energy $F$ of the topological string
in terms of the number $N_{\cR,Q,s}$ of BPS domain walls
of spin $s \in \frac{1}{2}\bZ$, charge $Q\in H_2(X,L)$ and representation $\cR$:
$$F = \sum_{\cR,Q,s} N_{\cR,Q,s} \cdot i\sum_{n=1}^\infty \frac{1}{2n\sin(n\lambda/2)}e^{n(-t_Q + is\lambda)}{\rm Tr}_\cR\prod_{i=1}^{b_1(L)}V_i^n$$
Here $\lambda$ is the string coupling and
$t_Q = \int_Q \omega,$ where $\omega$ is the symplectic form of $X.$ Also, implicit above is that we have chosen a basis $\{\gamma_i\}$ for $H_1(L,\bZ)$ and
have written $V_i$ for the monodromy of the flat connection on $L$ around $\gamma_i$.  Since we are interested
in the abelian theory on a single brane, we have $U(1)$ local systems and can ignore the
${\rm Tr}$ symbol, simply thinking of $V_i$ as a complex number of modulus one.

We are interested in the case where $X = \bC^3$, where the boundary map $\partial$ gives an isomorphism $H_2(X,L)\cong H_1(L)$
and the symplectic form is exact:  $\omega = d\theta.$  Then 
$t_Q = \int_Q \omega = \oint_{\partial Q}\theta$, and this quantitiy is a measure of the nonexactness
of $L$ through the periods of the one-form along $\partial Q \in H_1(L).$  These (real) periods vary as we move $L$ in the space of nonexact Lagrangians up to
Hamiltonian diffeomorhism.  They can be combined with the holonomies to form the complex open parameter
$x_i =  e^{\oint_{\partial Q}\theta} V_i,$ \`a la polar coordinates.  Note $x_i\in \bC^*.$
Let us put $q = e^{i\lambda}$.  We thus have
\begin{equation}
\label{eq:ov}
F =
\sum_{Q,s} \cN_{Q,s}\cdot - \sum_{n=1}^\infty \frac{1}{n(q^{n/2}-q^{-n/2})} (q^sx)^n
\end{equation}
Let us define the wavefunction $\Psi_{GW}$ to be $e^F.$
We will want to square the Ooguri-Vafa form of $e^F$ with the DT series of a quiver $\cQ$, to compare.\footnote{We use the script $\cQ$ because
the plain $Q$ already appeared
Ooguri-Vafa's formula, Equation \eqref{eq:ov}.}

We do this in Section \ref{sec:framing} after discussing some contexts where related results were derived earlier.

\section{Important Other Work}

All of this material has been studied before, though the explicit conjectures in open Gromov-Witten theory made in \cite{TZ,SZ} may be new.
Let us briefly review some of these other works.

\subsection*{The Abelian Theory on the Spectral Brane}

In this case, one again considers an M5-brane filling a three-dimensional spacetime
cross a three-fold in ${\bC^3}$, only now one looks at the effective theory on the brane.  In fact we take it one step further.
A general tactic when considering a brane supported on a space of the form $A\times B$ is to imagine
that one of the two factors is small and to consider the effective theory on the other factor.  In our case, $A$ is three- or four-dimensional
spacetime and $B$ is our Lagrangian $L$.
If you are counting protected quantities such as aspects of the BPS spectrum, you may investigate either limit.  This
set-up is explored in \cite{CEHRV,DGGo}.

The theory on $L$ is an abelian Chern-Simons theory, with asymptotic
boundary conditions defined by a flat connection on the boundary surface $\Lambda = \partial L.$
On general grounds, the partition
function of a theory with boundary defines a state in the Hilbert space of the boundary theory:  here we get a state in the quantization of the space $\cP_\Lambda$
of flat abelian connections on
the surface $\Lambda.$\footnote{Mathematicians know this from the modern definition of topological field theory.
Physicists know it in the following form.  Suppose $M$ is a manifold with boundary $\partial M = B$ and we investigate a field theory with fields $\varphi$.
The Hilbert space of this theory is comprised of functions of all boundary conditions $\varphi_B.$  The path integral
$Z(\varphi_B) = \int_{\varphi\vert_B = \varphi_B} e^{iS(\varphi)/\hbar}$ is just such a function.}
In fact, $L$ itself defines a Lagrangian submanifold $\cM_\Lambda$ of $\cP_\Lambda$ --- the connections which extend into $L$ --- and so the path integral determines a wavefunction
$\Psi$ on this Lagrangian subspace.

\subsection*{The Nonablian Theory on the base}

Our Lagrangian threefolds are embedded (or potentially immersed) in the cotangent $T^*B$ of a three-ball $B$, and are branched double covers over $B$, branched along a tangle as in \cite{CEHRV}.
If one lets the two sheets collide, one arrives at a nonabelian $GL_2$ theory on the ball.  
This interpretation is a special case of the more detailed exposition of \cite{DGGo} which considered more general three-manifolds with hyperbolic structures, especially knot complements, with gauge group $GL_K.$
They showed how different ideal triangulations of the hyperbolic manifold lead to mirror dualities in the effective three-dimensional theory.
They also show how patching moduli of the theory together create a global Lagrangian moduli space of the quantum cluster variety of decorated local systems.

\subsection*{The theory in three dimensions}

In \cite{CCV,CEHRV}, the authors understand the partition function of the three-dimensional theory as a generating function of BPS states of a four-dimensional theory.  To do so, they
considered the compactification from four to three dimensions wherein the moduli of the four-dimensional theory (equivalently of its Seiberg-Witten curve) vary in a prescribed way
so that all possible phases are sampled.\footnote{The
prescription is the horocycle flow. 
The periods of the Seiberg-Witten curve $\Sigma$ can be plotted as vertices of a polygon in the plane that gives a description of $\Sigma$ as a translation surface.
$SL_2(\bR)$ acts on the set of such surfaces.  In this language, the flow of \cite{CCV} is by the parabolic subgroup $\binom{1\;t}{0\;1}$.  This is called the horocycle flow; it preserves the hyperelliptic locus.}
They also use supersymmetric localization results to compute the partition function on the three-dimensional spacetime theory explicitly, as well as 1) relating it to a quantum-mechanical problem with
wavefunction $\Psi$ (same $\Psi$ as above) depending on moduli, 2) determining how $\Psi$ transforms under symplectic transformations of the moduli, and 3) relating this to braidings of the
branching tangle defining the compactifying Lagrangian.

%
%
%
%
%
%
%
%
%

\section{BPS Quivers, the KS COHA and DT Series}
\label{sec:framing}

To count BPS states of a given charge, one takes the dimension of the cohomology of the space of semiclassical BPS solutions.\footnote{This yoga may originate with
Gauntlett's work \cite{Ga} using Manton's low-energy analysis of monopos and Witten's description of vacua of supersymmetric quantum mechanics as cohomology classes.}
In our IIA description (``compactification'' on $\bC^3$), we have a four-dimensional $\cN=2$ supersymmetric theory.\footnote{Not really, since $\bC^3$ breaks no symmetry.
We imagine the $\bC^3$ as arising in some limit of a compact theory.  Alternatively, we can look for half-BPS states of a theory with more extended supersymmetry.}
We look for BPS states defined by branes wrapping internal dimensions of $\bC^3$ --- but since $\bC^3$ has no topology we will require a form of compactification
to make sense of the problem.  Requiring branes to end on a Legendrian surface partly does the trick.
%
%
The rest of the compactification data is what we call a ``framing.'' We imagine ``capping'' the surface at infinity with fixed handlebodies, glued at their boundary surface to $\Lambda.$
For the case where $\Lambda$ is a two-torus, this is explained in Section 5.1.1 of \cite{CEHRV}, where the gluing data is defined by a choice of $SL(2,\bZ)$ matrix.  
The different such framings, upon gluing to a Lagrangian filling, generate different compact three-folds with different intersection data.  Through matching calculations,
we understand this data to define the following BPS quiver.\footnote{The method of BPS quivers is outlined beautifully in Section 2 of \cite{ACCERV}.}

We focus on the case where $\Lambda$ is a particular genus-$g$ Legendrian surface defined in Section \ref{sec:framing}.
Fix a ``phase,'' namely a choice of Lagrangian filling $L$ (e.g. one coming from the connect sum of solid tori), which determines an isotropic (with respect to the
intersection form) subspace of $H_1(\Lambda,\bZ)$
which is killed upon inclusion into $L$ --- and then the choice of framing is a transverse isotropic subspace, and these are indexed by symmetric $g\times g$ matrix $A.$
Let $\cQ$ be the symmetric quiver with $g$ nodes and $A_{i,j} = A_{j,i}$ arrows betwen nodes $i$ and $j$.

Kontsevich-Soibelman \cite{KS} defined their Cohomological Hall Algebra (COHA) as a way of capturing
Harvey-Moore's construction \cite{HM} of the BPS algebra as encoded
by a BPS quiver, $\cQ.$\footnote{Given a BPS quiver $\cQ,$ the space of vacua of the corresponding quantum-mechanical problem is the space of representations of the
quiver, modulo gauge.  The collective coordinate method implies that the quantum states are given as cohomology classes of this space, understood to mean equivariant
classes with respect to the gauge group.  In some more detail, if $\cQ$ has $A_{i,j}$ arrows between nodes $i$ and $j$, then a dimension vector
$d\in {\bZ_{\geq 0}\setminus \{0\}}$
defines a moduli space $\bigoplus_{i,j=1}^n \bC^{A_{i,j} d_i d_j}/\prod_{i=1}^n GL(d_i)$}
Neglecting the algebraic structure, one gets a Poincar\'e polynomial $\Psi_{DT}$ counting BPS states, the so-called DT series of $\cQ$. 
For symmetric quivers without superpotential, the quiver varieties are contractible, with the only contributions coming from the equivariant cohomology
of the gauge group:  symmetric polynomials in several variables; their generating function counts Young diagrams and this is where
quantum dilogarithms arise (see Section
2 of \cite{KS}).
This DT series $\Psi_{DT}$ is a well-defined generating function of
\emph{all} BPS states, not just the stable ones, and as such depends on no notion of stability.  For non-symmetric quivers, a notion
of stability allows one to factorize $\Psi_{DT}$
as a product of quantum dilogarithms $\Phi(x) = \prod_{n=0}^\infty \frac{1}{1-q^nx}$ with exponents which define the stable BPS spectrum:  symbolically, $\Psi_{DT} = \Phi_1^{n_1}\Phi_2^{n_2}....$ 
Different notions of stability generate different factorizations, but the well-definedness of $\Psi_{DT},$ independent of the factorization, is a consequence of the cluster structure of the
cluster variety with seed $\cQ.$\footnote{See Zagier's paper for an excellent survey of the dilogarithm \cite{Z}, especially Section II.1.D.  The most basic identity behind the independence-of-factorization result is as follows.  Define
the (non-compact) quantum dilogarithm $\Phi(x) = \prod_{n=0}^\infty \frac{1}{1-q^nx}.$
Then if $yx = qxy$ we have $\Phi(x) \Phi(y) = \Phi(y) \Phi(-xy) \Phi(x)$.}

As stated above, we take $\cQ$ to be a symmetric quiver with a $g\times g$ symmetric adjacency matrix.
Kontsevich-Soibelman define the notion of admissibility for a DT series $\Psi_{DT}$,\footnote{See the product expansion of $F$
under Definition 6.2 of \cite{KS}, p. 305, for the $g=1$ case; our $N(d,s)$ is their $-c(n,i),$ where $d=n$ and $s=i$.} and prove it is
written as a product of dilogarithms to integral powers:
\begin{equation}
\label{eq:ks}
\Psi_{DT} = \prod_{d\in \bZ_{\geq 0}\setminus \{0\} } \prod_{s \in \frac{1}{2}\bZ} \Phi(q^s x^d)^{N(d,s)},
\end{equation}
where $x^d = x_1^{d_1}\cdots x_g^{d_g}$ and $N(d,s)\in \bZ.$

Now we can recognize that $Q$ and $d$ parametrize isomorphic quantities.  To compare $\Psi_{DT}$ and $\Psi_{GW}=e^F$, we identify $Q$ and $d$ and therefore
if $N(d,s)$ is to be compared with $\cN_{Q,s}$ (recall that $\cR$ has been fixed to the fundamental representation of $U(1)$), we should inspect $\log\Psi_{DT}$
We compute from Equation \eqref{eq:ks}:
\begin{align*}
\log \Psi_{DT} = \sum_{d\in \bZ_{\geq 0}\setminus \{0\}} \sum_{s \in \frac{1}{2}\bZ} N(d,s) &\cdot \log(\Phi(q^sx^d))\\
 {\rm "} \qquad\qquad &\cdot \sum_{m=0}^\infty \log(1-q^m(q^sx^d))\\
 {\rm "} \qquad\qquad &\cdot \sum_{m=0}^\infty \sum_{n=1}^\infty q^{mn}(q^sx^d)^{n}/n\\
  {\rm "} \qquad\qquad &\cdot \sum_{n=1}^\infty \frac{(q^sx^d)^n}{n}\frac{1}{1-q^n}\\
 {\rm "} \qquad\qquad &\cdot -\sum_{n=1}^\infty \frac{ q^{n/2}}{n(q^{n/2}-q^{-n/2})}(q^sx^d)^n
\end{align*}
We could achieve equality with Equation \ref{eq:ov} if we mysteriously declare that $H_2(X,\bZ)$ is one-dimensional and
all classses $Q$ lie in $(1,d)$ under the isomorphism $H_2(X,L)\cong \bZ\oplus H_1(L).$  Further, we set the pairing
$\int_Q \omega$ to be equal to $-i\lambda/2 + \oint_{\partial Q}\theta$,
in other words $e^{-t_Q+is\lambda}\prod_i V_i = q^{1/2}(q^s x^d)$.
We have no good explanation for this, but note that a nearly identical
relationship between symplectic form and string coupling appears in Equation (3.8) of \cite{ANV} with the remark
that it is not unfamiliar.

Under the WKB approximation,
we want to write $\Psi = e^F,$ with leading behavior $F = W/\hbar.$   Here $\hbar = \lambda = \log(q),$ and
the Ooguri-Vafa $1/d^2$ genus-zero disk multiple-cover formula corresponds to the fact that $\lim_{\hbar\to 0} \hbar\log \Phi(x) = \Li_2(x).$

The upshot of all this is that, modulo a mysterious symplectic form, all-genus Ooguri-Vafa integrality agrees with integrality of the full DT series, and the definitions align
for string compactifications on $\bC^3$ for unstacked (rank-one) branes.\footnote{The relationship between the quantum dilogarithm and
open-string calculations was found in \cite{ADKMV}, though a direct connection to Ooguri-Vafa integrality does not seem to have been articulated.}

The framing duality conjecture \cite{SZ} asserts the above, along with a recipe for identifying which branes belong to which under this identification.  
We explain which branes --- \emph{and with which framings} --- to associate to which quivers.

\section{Framing Duality}

In \cite{ADKMV}, one of the main examples studied was the IIB string theory on the mirror of $\bC^3,$ i.e. the space $\{wz = x + y - 1\} \subset \bC^2\times (\bC^*)^2.$
Within this background, the authors studied B-branes which are a copy of $\bC$ and are defined by the equation $w=0$ with $x$ and $y$ constant.  The moduli space of such branes is $\{x+y-1\}\subset (\bC^*)^2,$
a pair of pants.  The mirror of these branes are the Harvey-Lawson special Lagrangians\footnote{Defined by $|z_1|^2-\epsilon^2 = |z_2|^2 = |z_3|^2,$
${\rm Arg}(z_1z_2z_3)=0$, for example.  Others are given by permuting $(1,2,3).$}
in $\bC^3$ studied by Aganagic-Vafa.\footnote{One
could also consider the branes defined by the equation $z=0$, with $x$ and $y$ constant.  This suggests that the true moduli space of branes is two disjoint copies of
the pair of pants, and for each brane in one component, there is a unique brane in the other component that intersects it.  The intersection is not transverse.  Mirror to these we have
the branes in $\bC^3$ defined by the same geometric condition as the Aganagic-Vafa branes, but instead of the usual condition ${\rm Arg}(xyz)=0$ we can impose
${\rm Arg}(xyz)=\pi.$  This family is then special Lagrangian with phase shifted by $1$, suggesting an anti-brane interpretation.  This relationship is detailed in the last
paragraph of Section 2 of \cite{ADKMV}.
In \cite{SZ} we find in fact two disjoint copies of the pair of pants.  To study the usual Aganagic-Vafa branes we only considered one copy, but both are physical.
Similarly, other components appear for the generalizations of the Aganagic-Vafa branes discussed here.}

In \cite{ADKMV} the authors also consider multiple branes which are $g$ disjoint copies of $\bC$ labeled by $g$ points on the mirror curve $\{x+y=1\}\subset (\bC^*)^2.$
One possible mirror for such branes would be $g$ disjoint Aganagic-Vafa branes.  However, it would seem that framings of such branes would be given
by $g$ integer values.  Instead, we propose that the mirrors are Lagrangians with $b_1(L) = g$ that are asymptotic to certain genus-$g$ Legendrian surfaces $\Lambda_g$, which we define below.
These genus-$g$ Legendrians are connect sums of the tori at the boundary of Aganagic-Vafa branes and as such
have the same moduli ($g$ copies of the pair of
pants), but they have more framings (which allow the connection to DT quivers) and moreover fit into a larger family of branes related through cluster theory.

We first recall the general construction of a Legendrian submanifold of $T^*M\times \bR$ by an immersed, transverse-to-vertical hypersurface in $M\times \bR$.  Coordinatize the $M$ factor with $x = (x^i)$ and the $\bR$ factor with $z$.  Then over a patch, components of the hypersurface, being non-vertical, can be locally defined by a function $z(x).$  Now simply define $y_i = \frac{\partial z}{\partial x^i}$.  The hypersurface $H$ is the (front)
projection of the Legendrian $\Lambda_H \to M\times \bR$.  Now set $M= S^2.$ We will take $H$ to be a two-sheeted cover over $S^2$ with sheets crossing over
$\Gamma\subset S^2$ a cubic graph.  Over the edges of $\Gamma,$ the sheets cross.  Over the vertices, the different sheets become parallel so that $H$ is not immersed.  Nevertheless, the two sheets are set to be parallel there and so $\Lambda\to \Gamma$ makes sense as a branched double cover over $S^2,$ branched over the vertices.  
A genus-$g$ surface is defined by a graph $2g+2$ vertices.
\vskip-.1in
$$
\begin{tikzpicture}
\node at (0,0) {Vertex of $\Gamma$};
\node at (0,-1.5) {$\phantom{a}$};
\end{tikzpicture}
\quad
\begin{tikzpicture}
\draw[red,thick] (-1,-.13) -- (0,0);
\draw[red,thick] (.6,.6) -- (0,0);
\draw[red,thick] (.5,-.6) -- (0,0);
\node at (0,-1.5) {$\phantom{a}$};
\end{tikzpicture}
\quad
\qquad
\includegraphics[scale = .2]{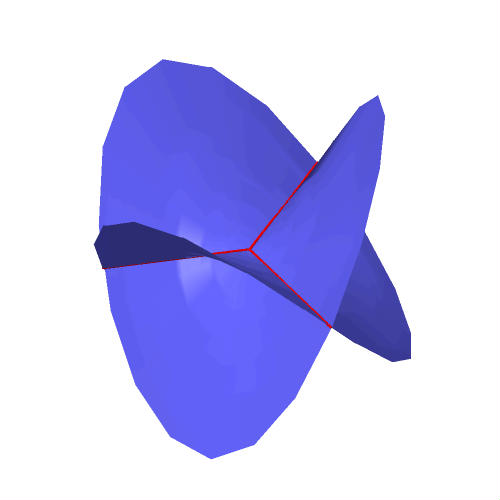}
\qquad
\begin{tikzpicture}
\node at (0,0) {Wavefront near a vertex};
\node at (0,-1.5) {$\phantom{a}$};
\end{tikzpicture}
$$

%
%
%
%
%
%

We define $\Lambda_g$ to be the Legendrian defined by the graph $\Gamma_g$,
a $g$-fold blow-up of the $\Theta$ graph.\footnote{Other graphs
are related by mutations and have wavefunctions that one can compute via cluster theory \cite{SZ} (see Section \ref{sec:computations}),
though we know of no BPS quiver to go with them.}
It looks like a canoe with $g$ seats:
\begin{center}
\begin{tikzpicture}[decoration=brace,scale =.4]
%
\node at (0,0) {$\bullet$};
\node at (1.2,.8) {$\bullet$};
\node at (1-.2,-.8) {$\bullet$};
\node at (2.2,.8) {$\bullet$};
\node at (3.2,.8) {$\bullet$};
\node at (4.2,.8) {$\bullet$};
\node at (5.2,.8) {$\bullet$};
\node at (6.2,.8) {$\bullet$};
\node at (2-.2,-.8) {$\bullet$};
\node at (3-.2,-.8) {$\bullet$};
\node at (4-.2,-.8) {$\bullet$};
\node at (5-.2,-.8) {$\bullet$};
\node at (6-.2,-.8) {$\bullet$};
\node at (7,0) {$\bullet$};
\node at (3.7,.8) {$\cdots$};
\node at (3.3,-.8) {$\cdots$};
\draw (0,0)--(1.2,.8)--(2.2,.8)--(3.2,.8)--(3-.2,-.8)--(2-.2,-.8)--(1-.2,-.8)--(0,0);
\draw (7,0)--(6.2,.8)--(5.2,.8)--(4.2,.8)--(4-.2,-.8)--(5-.2,-.8)--(6-.2,-.8)--(7,0);
\draw (1.2,.8)--(1-.2,-.8);
\draw (2.2,.8)--(2-.2,-.8);
\draw (3.2,.8)--(3-.2,-.8);
\draw (4.2,.8)--(4-.2,-.8);
\draw (5.2,.8)--(5-.2,-.8);
\draw (6.2,.8)--(6-.2,-.8);
\draw (0,0) .. controls (-.8,-2.8) and (8,-2.8) .. (7,0);
\node[blue,thick] at (3.5,-1.5) {$\Gamma_g$};
\draw[decorate, yshift=2ex]  (1.2,1) -- node[above=0.4ex] {$g$ seats}  (6.2,1);
\end{tikzpicture}
\end{center}
Up to issues with basepoints, this Legendrian surface has the same dga as the disjoint union of $g$ tori, whose solid tori fillings are the Aganagic-Vafa branes.

The ball $B$ has boundary $S^2$ and we consider a (non-exact) Lagrangian threefold $L\subset T^*B$ with boundary asymptotic to $\Lambda_g$, a branched double
cover over a tangle in $B$ ending on the vertices of $\Gamma_g$ \cite{TZ}.
The inclusion of the boundary $\Lambda_g\hookrightarrow L$ defines
a map on the first homology, and the kernel $K$ is an isotropic subspace of $H_1(\Lambda_g,\bZ)$ which we call a ``phase," following Aganagic-Vafa.\footnote{In their
example, there were three phases in $H_1(T^2,\bZ)\cong \bZ^2$ given by the spans of $(1,0), (0,1), (-1,-1)$.}
We call a ``framing'' a transverse isotropic subspace.  Given one phase and framing, the other framings are parametrized by an integral, symmetric matrix.  In fact, there is
a standard phase obtained by expressing the canoe graph as a mutation of a ``necklace'' graph, which has a canonical phase and framing. 
We fix this phase and label the framings by a $g\times g$ symmetric matrix $A.$  Then $L$ and this framing defines an open Gromov-Witten problem.\footnote{Jake Solomon and
Sara Tukachinski are developing the open Gromov-Witten theory we discuss here.}  

Now let $\Psi^A_{GW}(\Gamma_g)$ be the generating function of open Gromov-Witten invariants of $L$ in framing $A$.  Let $\cQ$ be the quiver with adjacency matrix $A$ and let
$\Psi^A_{DT}$ be the DT series for $\cQ$.  We conjecture
$$\Psi^A_{GW}(\Gamma_g) = \Psi^A_{DT}.$$

In fact, we can define conjectural $\Psi^A_{GW}(\Gamma)$ for any cubic planar graph $\Gamma$ in any phase or framing.

\section{Computing Wavefunctions}
\label{sec:computations}

Given a cubic planar graph $\Gamma$, we have defined (up to isotopy) a Legendrian surface $\Lambda_\Gamma.$  It defines
a moduli space $\cM_\Gamma$ of rank-one (unstacked) Lagrangian branes with asymptotic conditions defined by $\Lambda_\Gamma.$
The method of Aganagic-Vafa (which applies to the case where $\Gamma$ is the tetrahedron graph) is
to write $\cM_\Gamma$ as a subspace of an algebraic torus $\cP_\Gamma$:  in their example, this looks like $x+y=1$ inside ($\bC^*)^2.$  
Here we have implicitly chosen a phase and framing, which define coordinates on $\cP_\Gamma.$

Using the methods of sheaf theory, one can describe the moduli space of rank-one objects in such a Fukaya category:
in fact $\cM_\Gamma$ turns out to be the set of map colorings of $\Gamma$, with colors chosen from $\bP^1$, modulo $PGL_2$ \cite{TZ}.
The algebraic torus is the rank-one local systems on $\Lambda_\Gamma$, and this space quantizes to the quantum torus. 
For the Aganagic-Vafa example, the quantum algebra is $yx = qxy,$ where as before $q = e^\lambda.$  Thinking of $y$ as $e^{\lambda\partial_x},$
this algbra acts on functions as $(x\cdot f)(x) = xfx)$ and $(y\cdot f)(x) = f(qx).$
In fact, $\cM_\Gamma$
quantizes to a cyclic ideal for this algebra defined by $\Psi_\Gamma.$ 
In the Aganagic-Vafa case, $(y+x-1)\cdot \Psi_\Gamma = 0$ means $\Psi_\Gamma(qx) = (1-x)\Psi_\Gamma(x),$ giving $\Psi_\Gamma = \Phi,$
the Ooguri-Vafa result (modulo our funny symplectic form). 
This same structure was found and explored in detail in \cite{DGGo}, though not to describe Fukaya moduli. 

The cubic planar graphs that describe our Legendrians are dual to triangulations of the sphere, which label cluster charts.
The algebraic torus $\cP_\Gamma$ is just one chart in a larger space (bigger than the space of branes?) $\cP,$ a cluster variety.
In fact, the different $\cM_\Gamma$ glue together to form a single universal quantum Lagrangian $\cM\subset \cP$ \cite{DGGo,SZ},
meaning the cluster transformations also transform the equations defining the $\cM_\Gamma.$\footnote{In
\cite{CEHRV,DGGo}, mutations are shown to effect three-dimensional mirror symmetries of the spacetime theory.}
  
The good news is that phases and framings can be transferred with mutations as well.  Even better, cluster transformations $\mu:\Gamma\leadsto\Gamma'$ are
defined by conjugation with appropriate\footnote{Here ``appropriate'' means one has to choose signs and powers and arguments carefully.} dilogarithms $\Phi_\mu$.
But then we can conclude $\Psi_{\Gamma'} = \Phi_\mu\Psi_\Gamma.$
\begin{center}
\begin{tikzpicture}[scale=.9]
\newcommand*{\boff}{10}; \newcommand*{\aoff}{35}; \newcommand*{\rad}{3};
\coordinate (a) at (0,-1.4);
\coordinate (b) at (4,-2);
\coordinate (c) at (6,0);
\coordinate (d) at (2,2);
\coordinate (e) at (0,1);
\draw[thick] (a) to[out=300,in=180] (b) to[out=180+180,in=270] (c) to[out=90,in=0] (d) to[out=180,in=50] (e) to[out=230,in=120] (a);
\draw[thick,blue] (c) to[out=210,in=320] (e);
\node at (5,1.9) {$\mathcal P$};
\node at (.8,1) {$\mathcal M$};
\node at (4.5,.7) {$\mathcal P_\Gamma$};
\node at (4.2,-.3) {$\mathcal M_\Gamma$};
\draw[thick,red] (4-.5,-.5+.866) -- (4-.5,-.5-.866);
\draw[thick,red] (4-.866,-.5-.5) -- (4+.866,-.5-.5);
\node[red] at (2.8,.4) {phase};
\node[red] at (2.5,-1) {frame};
\draw[thick] (4,-.5) circle (1cm);
\end{tikzpicture}
\end{center}
This means we can compute conjectural Gromov-Witten invariants
of Lagrangian fillings, with any phase or framing, of any Legendrian surface with a cubic graph description!
Some examples were computed in \cite{TZ} and \cite{LZ,Zh}.  The computational schema described here is detailed in \cite{SZ}.

\vspace{2mm}\noindent {\bf Acknowledgements.}
I would like to thank David Treumann and Linhui Shen, my collaborators on the mathematical material on which this paper relies.
I thank Roger Casals and Melissa Liu for helpful conversations.
I would also like to thank Tudor Dimofte, Andy Neitzke and Cumrun Vafa for explaining relevant aspects of their work.
This work is supported by NSF grant DMS-1708503.

\end{document}